\documentclass{nature2}
\usepackage{graphicx}
\usepackage{tabularx}
\usepackage{longtable}
\title{An intuitive 3D map of the Galactic warp's precession traced by
  classical Cepheids}

\author{Xiaodian Chen$^{1}$, Shu Wang$^2$, Licai Deng$^{1,3}$, Richard de
  Grijs$^{3,4,5,6}$, Chao Liu$^1$ and Hao Tian$^1$}

\begin{document}

\maketitle

\begin{affiliations}
 \item Key Laboratory for Optical Astronomy, National Astronomical
   Observatories, Chinese Academy of Sciences, 20A Datun Road,
   Chaoyang District, Beijing 100012, China
 \item Kavli Institute for Astronomy and Astrophysics, Peking
   University, Yi He Yuan Lu 5, Hai Dian District, Beijing 100871,
   China
 \item Department of Astronomy, China West Normal University, Nanchong
   637009, China
 \item Department of Physics and Astronomy, Macquarie University,
   Balaclava Road, Sydney, NSW 2109, Australia
 \item Research Centre for Astronomy, Astrophysics and Astrophotonics,
   Macquarie University, Balaclava Road, Sydney, NSW 2109, Australia
 \item International Space Science Institute--Beijing, 1 Nanertiao,
   Zhongguancun, Hai Dian District, Beijing 100190, China
\end{affiliations}

\begin{abstract}
The Milky Way's neutral hydrogen (H{\sc i}) disk is warped and
flared\cite{Levine06, Kalberla07}. However, a dearth of accurate H{\sc
  i}-based distances has thus far prevented the development of an accurate
Galactic disk model. Moreover, the extent to which our Galaxy's
stellar and gas disk morphologies are mutually consistent is also
unclear. Classical Cepheids, primary distance indicators with distance
accuracies of 3--5\%\cite{{Wang18}}, offer a unique opportunity to
develop an intuitive and accurate three-dimensional picture. Here, we establish a
robust Galactic disk model based on 1339 classical Cepheids. We
provide strong evidence that the warp's line of nodes is not oriented
in the Galactic Center--Sun direction. Instead, it subtends a mean
angle of 17.5$^\circ \pm 1^\circ$ (formal) $\pm 3^\circ$ (systematic)
and exhibits a leading spiral pattern. Our Galaxy thus follows Briggs'
rule for spiral galaxies\cite{Briggs90}, which suggests that the
origin of the warp is associated with torques forced by the massive
inner disk\cite{Shen06}. The stellar disk traced by Cepheids follows
the gas disk in terms of their amplitudes; the stellar disk extends to
at least 20 kpc\cite{Liu17,Wangh18}. This morphology provides a
crucial, updated map for studies of the kinematics and archaeology of
the Galactic disk.
\end{abstract}

 We have compiled samples of classical Cepheids from the
  Wide-field Infrared Survey Explorer (WISE) catalogue of periodic
  variables\cite{Chen18a} (our `WISE Cepheid sample') as well as from a number
  of optical surveys (collectively referred to as our `optical Cepheid 
  sample'). We will discuss both samples separately, because the
  catalogues' optical and infrared passbands are characterised by
  significantly different photometric and extinction sensitivities.
Highly accurate Cepheid distances can be estimated using their
well-established wavelength-dependent period--luminosity relations. To
mitigate the influence of extinction in the Galactic plane and of
photometric uncertainties at infrared wavelengths, we adopted the
`infrared multi-passband optimal distance method'\cite{Chen18b} 
  to determine accurate Cepheid distances. Contaminants, including
Type-II Cepheids, long-period eclipsing binaries and quasi-periodic
variables  were removed using {\sl Gaia} Data Release 2
parallaxes\cite{Gaia18}. Cepheids located in areas centered on the
Magellanic Clouds were also excluded. Careful
sample selection resulted in a tally of 2330 classical Cepheids for further analysis.

\begin{table}
\vspace{-0.0in}
\begin{center}
\caption{\label{t1}Parameters of the linear and power-law warp
  models applied. RMSE: Root mean square error.}
\vspace{0.15in}
\begin{tabular}{lccccc}
\hline
\hline
                                                                                                            
              \multicolumn{5}{c}{$z_{\rm w}=a(R-R_{\rm w})^b\sin(\phi-\phi_{\rm w})$}  \\ 
\hline                               
          & $R_{\rm w}$ (kpc) &$\phi_{\rm w}$ ($^\circ$)  & $a$  &	$b=1$  &  RMSE \\         
All      & $ 9.26\pm0.19$&$17.4\pm1.2$&$0.148\pm0.006$  & &0.256\\                               
WISE & $ 9.31\pm0.20$&$18.0\pm1.6$&$0.146\pm0.006$  & &0.215\\                               
Optical  & $9.01\pm0.40$&$16.1\pm1.7$&$0.148\pm0.011$  & &0.292\\                               
\hline                                                                                    
           & $R_{\rm w}$ (kpc) &$\phi_{\rm w}$ ($^\circ$)  & $a$  &	$b$	 &  RMSE   \\            
All      & $7.72\pm0.27$&$17.5\pm1.0$&$0.060\pm0.012$  &$1.33\pm0.08$ & 0.210\\                
WISE & $6.93\pm0.72$&$18.6\pm1.4$&$0.028\pm0.015$  &$1.61\pm0.19$& 0.188\\                
Optical  & $7.85\pm0.27$&$15.8\pm1.3$&$0.093\pm0.019$  &$1.14\pm0.09$ &0.225\\                
\hline
\end{tabular}
\end{center}
\end{table}

Distances were converted to 3D $XYz$ and spherical $R \phi z$
coordinates by adopting a reference frame centered on the Galactic
Center and a solar Galactocentric distance $R_0=8.0$ kpc. Here, $\phi$
is the Galactocentric angle in the anticlockwise direction (aligned
with the disk's rotation axis) with respect to the solar position
($\phi_{\odot}=0^\circ$). Since {\sl Gaia} parallaxes are
  reliable within $\sim$5 kpc, we only selected Cepheids within the
volume $R<20, |z|<2$ kpc to avoid significant contamination by Type-II
Cepheids. Our downselection included 1459 Cepheids with distances
accurate to $<$5\%, corresponding to a distance modulus standard
deviation $<$0.108 mag. Cepheids located clearly away from the
best-fitting warp model ($\Delta>1$ kpc) were also excluded (see
below). Our final sample contained 1339 Cepheids.

\begin{figure}
\begin{center}
\includegraphics[width=\textwidth]{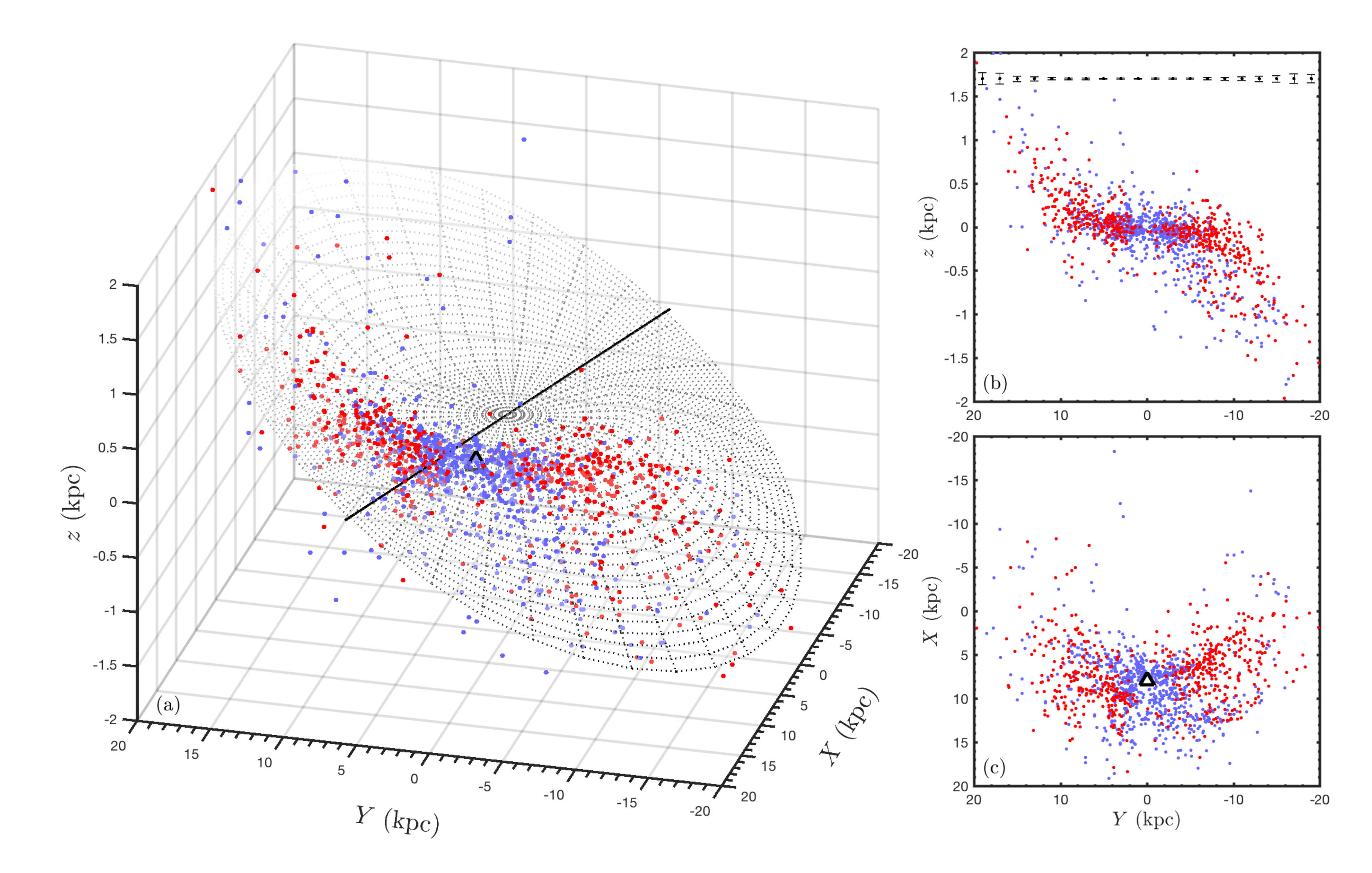}
\caption{\label{f1}{\bf 3D map of the Milky Way disk traced by
    Cepheids.} a: Red and blue dots represent, respectively, 585
    and 744 Cepheids discovered in infrared (WISE) and optical
  passbands. The black upward-pointing triangle is the position of the
  Sun. Warp features are seen down to negative $z$ on the right and up
  to positive $z$ on the left. The grid is our best model of a
  power-law warped disk (see Table 1) and the black solid line denotes
  the LON, $\phi=17.5^\circ$. The LON obviously deviates from the
  Sun--Galactic Center direction (see the online video for a better
  impression). Projections onto the $Yz, XY$ plane are shown in panels
  b and c; $z$-height error bars are included for different
    values of $Y$.}
\end{center}
\end{figure}

Figure 1a shows the 3D distribution of our final Cepheid sample. It covers two-thirds of
the disk. The bluish-violet `optical' Cepheids are distributed in the
solar neighbourhood and the Galactic anticenter direction. The red
WISE Cepheids are predominantly located on either side of our optical
sample. The Galactic warp is clearly visible, particularly its
downward deviation from the Galactic plane. To model the warp, we first adopted
the commonly used model, $z_{\rm w}=a(R-R_{\rm w})^b\sin(\phi-\phi_{\rm
  w})$, where $z_{\rm w}$, $R_{\rm w}$ and $\phi_{\rm w}$ are the warp
height, its onset radius and the line of nodes (LON), respectively. Linear and power-law nonlinear least-squares fits were applied to all sample Cepheids. For $R<9.0$ kpc, a power-law description is most
appropriate (see the grid in Figure \ref{f1}), whereas the
distribution becomes linear for $R>9.0$ kpc: see Figure \ref{extend1}
and Table \ref{t1}. In addition, our WISE Cepheid sample is contaminated by
fewer overtone Cepheids than the optical sample (see below). To
limit the impact of possible selection effects, we fitted our model to the WISE
and optical Cepheid samples both separately and simultaneously. Owing to their
more extended distribution across the disk, parameters based on the WISE
Cepheids are intrinsically more accurate. Nevertheless, all model
results are internally consistent given the uncertainties. This implies that selection
effects are thus minor or indeed negligible.

We carefully checked our results based on Markov-chain Monte Carlo
(MCMC) simulations using the MATLAB MCMC toolbox\cite{Haario06}. The
resultant excellent agreement, both as regards the parameter values
and the well-defined Gaussian distributions (see Figure \ref{extend3})
implies that our results are indeed robust. The LON angle $\phi_{\rm w}$ is
insensitive to the adopted type of model because of the weak
correlation between $\phi_{\rm w}$ and $(a, b, R_{\rm w})$. We also
considered a warp model expressed in spherical
coordinates\cite{Abedi14}, but this only marginally affected the LON
angle, because our Cepheids are located close to the Galactic plane.

\begin{figure}
\begin{center}
\includegraphics[width=140mm]{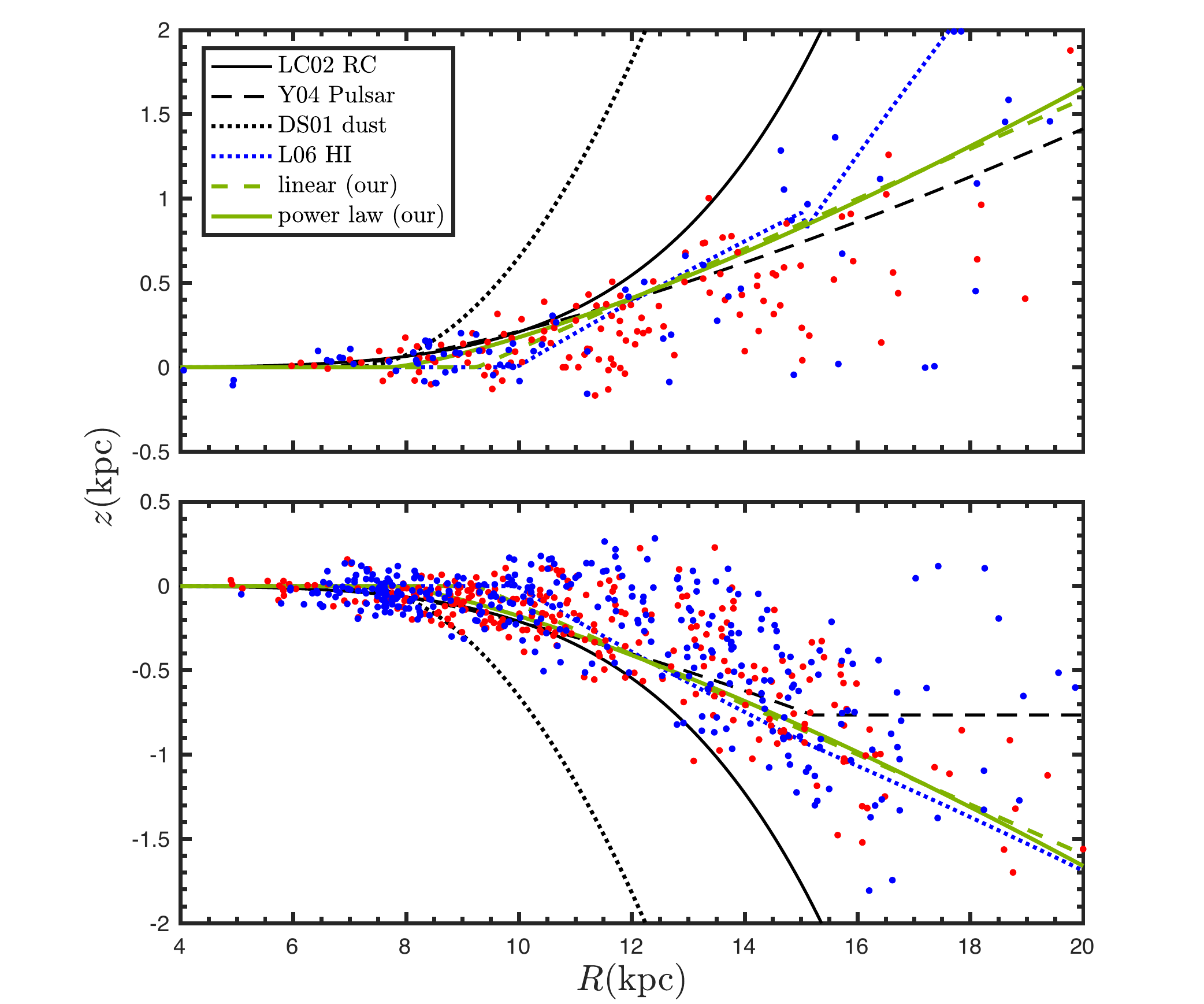}
\caption{\label{f2}{\bf Maximum $z$ heights of the warps.} The top and
  bottom panels represent the northern and southern warps,
  respectively. Red and blue dots represent Cepheids found in,
  respectively, infrared (WISE) and optical passbands. The green solid and dashed lines are Cepheid warps derived in this study based on the power-law and linear warp models, respectively. Comparison with
  other warp determinations; LC02: red clump (RC) giants
  warp\cite{Lopez02}; Y04: pulsar warp\cite{Yusifov04}; DS01: dust
  warp\cite{Drimmel01}; L06: H{\sc i} warp\cite{Levine06}.}
\end{center}
\end{figure}

To explore whether our derived stellar warp morphology agrees with
that of the Galaxy's H{\sc i} gas warp, at least regarding their $z$
heights, we projected the Cepheids' 3D distribution onto the plane
defined by the maximum height of the warp, assuming
$\sin(\phi-\phi_{\rm w})=1$. Figure \ref{f2} compares our model with
both the H{\sc i} warp and the distributions defined by a number of
other tracers\cite{Levine06,Drimmel01,Lopez02,Yusifov04}. The Cepheid
warp is in excellent agreement with the H{\sc i} warp out to $R
\simeq$ 15 kpc ($\Delta |z|<50$ pc). At larger radii, the H{\sc i} warp's Fourier m = 0 and m = 2 modes start to dominate. This
enhances its amplitude, particularly in the northern warp. Despite the
detection of a similar $m=2$ amplitude for our Cepheid sample,
$W_2=0.14\pm0.03$ kpc kpc$^{-1}$, that sample's rather different LON
angle $\phi_{\rm W2}=174\pm10^\circ$---which deviates from the H{\sc
  i} LON by $\sim$$45^\circ$---precludes us from assessing this type
of behaviour in the Cepheid warp (see Figure \ref{extend3}).

Although the warp traced by pulsars is generally comparable with our
Cepheid warp, the roughly 30\% uncertainties affecting pulsar
distances render any details unreliable. The warps traced by Two
Micron All Sky Survey\cite{Skrutskie06} (2MASS) red clump stars agree
with neither the Cepheid nor the H{\sc i} warps. This may be owing to
sampling incompleteness at $R>13$ kpc: since red clump stars are some
four magnitudes fainter than Cepheids, stellar crowding, the
  Sun's location close to the warp and background glare towards the
  Galactic Center imply that the former are more likely detected
close to the Galactic anticenter. This problem is compounded if the
warp's LON deviates from the Sun--Galactic Center direction (see
below). Evolution of the warp's morphology was initially suggested
based on the shallow 2MASS data and a population synthesis
model\cite{Amores17}. However, the recent warp kinematics
map\cite{Poggio18} based on {\sl Gaia} proper motions, combined with a
flat rotation curve, suggests that the old and young stellar
populations exhibit similar warp features.

The warp’s LON, combined with theoretical predictions, can help us constrain the warp’s
origin. However, the Milky
Way's LON has never been studied using tracers with distance
accuracies better than 20\%. The statistical and systematic
uncertainties in the distances unavoidably caused reduced
  accuracies in both the LON's mean value and its shape. The LON of
the Milky Way's H{\sc i} warp is closely aligned with the Galactic
Center--Sun direction\cite{Burton88}, $\phi_{\rm
  w}=0^\circ$. Similarly, observations of dust and stars with the Cosmic Background Explorer (COBE)\cite{Drimmel01} and of red clump stars with 2MASS\cite{Lopez02}
yielded $\phi_{\rm w}=0^\circ$ and $\phi_{\rm w}=-5\pm5^\circ$ (statistical error),
respectively. Although a 2MASS-based study of red clump stars and red
giants\cite{Momany06} found $\phi_{\rm w}\sim15^\circ$, its authors
did not provide an assessment of the uncertainties nor of their
selection effects. Our application of accurate distances
implies that the Milky Way's global LON deviates significantly from
the Sun--Galactic Center direction, $\phi=17.5^\circ \pm 1.0^\circ$
(formal, including propagation errors) $\pm 3.0^\circ$. The latter,
systematic error is introduced by the combination of the Sun's height
above the Galactic midplane, $z=25$ pc\cite{Bland-Hawthorn16}, and the
difference between the current-best Galactocentric distance,
$R_\odot=8.3$ kpc\cite{deGrijs16}, and that adopted here. Note that we determined the mean LON angle and its uncertainty assuming that the LON does not vary with Galactocentric radius.
  
\begin{figure}
\begin{center}
\includegraphics[width=\textwidth]{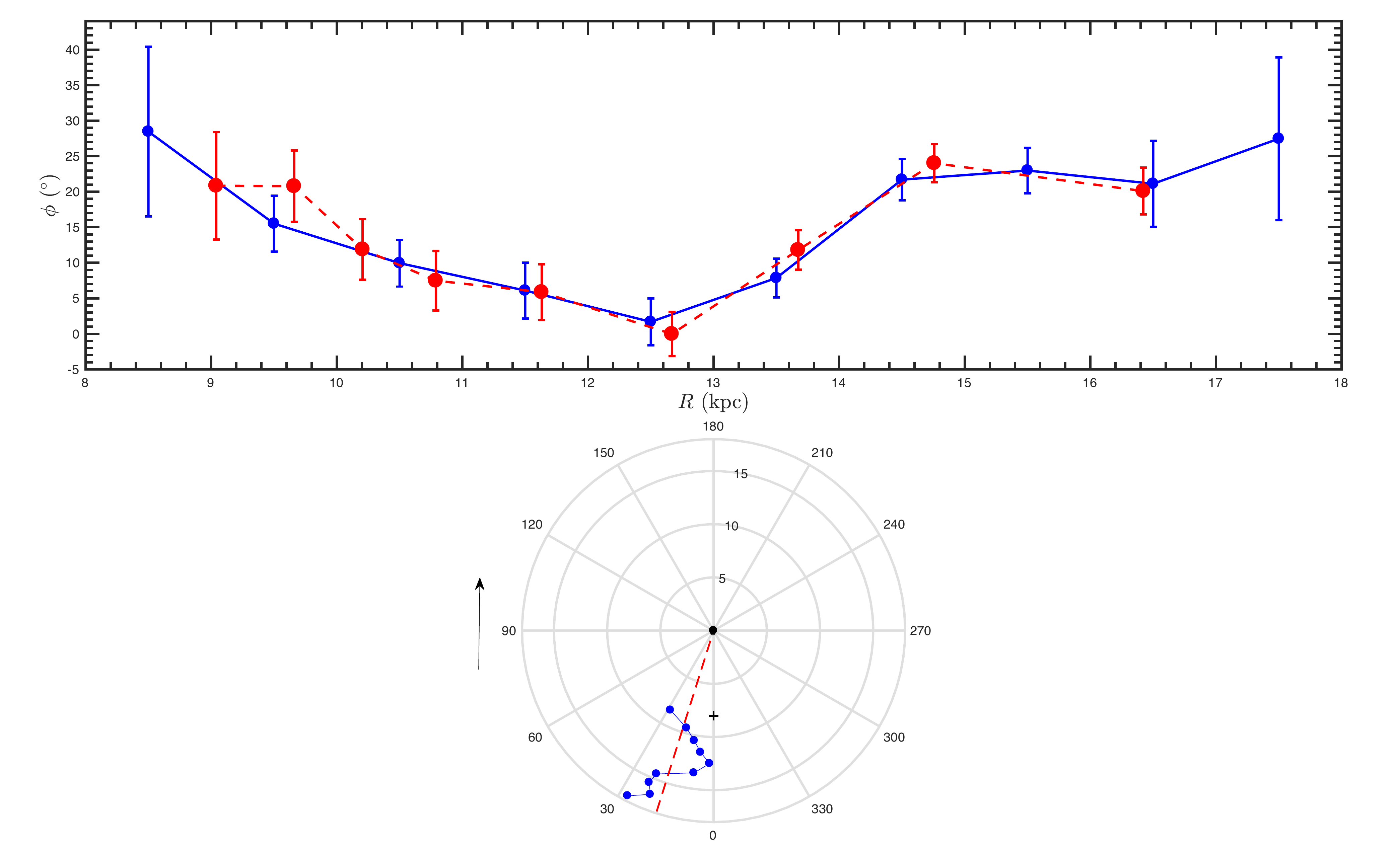}
\caption{\label{f3}{\bf The Milky Way's line of nodes.} Top: Variation
  of the warp's LON ($\phi$) with galactocentric radius. Blue and red
  dots and error bars denote $\phi$ determined on the basis of two samples:
  (i) Cepheids in bins of $R_i-0.5<R<R_i+0.5$ kpc and (ii) identical
  numbers of Cepheids in each bin ($R_i$ is the variable
  galactocentric radius). All samples show that the LON increases for
  $R = 12$--15 kpc. Bottom: LON in polar coordinates. The arrow
  denotes the direction of rotation of the Milky Way's disk.}
\end{center}
\end{figure}

To ascertain whether the LON is stable as a function of galactocentric radius, we subdivided our Cepheid samples using two selection cuts: (i) objects within 1.0 kpc-wide bins and (ii) equal
numbers of Cepheids (95) in each bin; 1.0 kpc is the optimal bin size
according to the Freedman--Diaconis rule\cite{Freedman81}. The
power-law warp model was fitted to both samples, adopting fixed
$(R_{\rm w}, a, b)$. We also estimated the LONs and their
uncertainties: see Figure \ref{f3} (top). Both trends are globally
similar for $R=9$--16 kpc, which thus suggests that the general trend
is not affected by problems associated with insufficient sample
sizes. The blue dots are also plotted in polar coordinates to allow
comparison with the disk's rotation (see Figure \ref{f3}, bottom). To validate the derived LON, we
performed MCMC simulations of $\phi$ for different conditions. They
included fixed and free $(a, b, R_{\rm w})$ parameters, 5\% and 10\%
limiting distance accuracy cuts and resampling of the Cepheids in the
northern and southern warps. We also estimated the propagated
uncertainties associated with the objects' distances. All resulting
LONs are mutually consistent (see Figure \ref{extend4}). We adopted
the largest values of the statistical and systematic errors as our final
uncertainties.

We also attempted a kinematic analysis of the Cepheid sample, adopting
the proper motions and radial velocities from {\sl Gaia} Data Release
2. Since two-thirds of the Cepheids do not have radial velocity
measurements, we evaluated their radial velocities assuming a flat
rotation curve\cite{Reid14}, $v_{\rm c} =240$ km s$^{-1}$. The typical
uncertainty was based on the scatter in the rotation velocities of the
other one-third of our sample with such measurements, $\Delta v_{\rm
  c} = 13$ km s$^{-1}$. We converted the 3D velocities to $(v_r,
v_\phi, v_z)$ in Galactocentric coordinates,
assuming\cite{Schonrich10} $(U, V, W) = (11.1, 12.24, 7.25)$ km
s$^{-1}$. The mean uncertainty $\langle \sigma_{v_z} \rangle$ is
  around 4.2 km s$^{-1}$; only Cepheids with uncertainties
$\sigma_{v_z}<5$ km s$^{-1}$ are plotted in the $XY$--$v_z$
diagram. The spatial LON agrees globally with the $v_z$ maxima distribution.

A clear increase of the LON is apparent at $12 \le R \le 15$ kpc in
Figure \ref{f3} (top). In theory\cite{Shen06}, the retrograde
precession rates of the outer disk caused by, respectively, the massive inner
disk and external torques scale as $R^{-4}$ and $R$. This represents
the first clue that the Galactic warp traces a leading spiral pattern, which validates the notion that the origin of the outer disk’s pattern is predominately induced by torques associated with the massive inner disk. However,
near $R = 15.5$ kpc the LON appears to twist, possibly because of
either external forcing of the misaligned outer halo or satellite
accretion. In addition, within $R=12$ kpc the decrease of the LON with
radius is likely caused by a decrease of the rotational speed (see the
diagonal ridge in the $v_\phi-R$ diagram\cite{Antoja18}).

Although LON precession of the Galactic warp has not yet been
detected, it has been reported for 12 other spiral
galaxies\cite{Briggs90}. Those latter galaxies approximately follow
Briggs' rule: the LON remains straight within $R_{25}$ and advances in
the rotation direction from around the Holmberg radius, $R_{\rm
  {Ho}}$. The LON traced by Cepheids conforms with this
rule. Quantitatively, for a radial thin-disk scalelength of $R_{\rm
  d}=2.6\pm0.5$ kpc\cite{Bland-Hawthorn16}, the Milky Way's
$R_{25}=3.0R_{\rm d}$ and $R_{\rm{Ho}}=4.4R_{\rm d}$\cite{Freeman70}
are located at 7.8 and 11.4 kpc, respectively. These radii agree well
with the onset radius of the warp and the leading spiral
pattern. Although the uncertainty in the scalelength is significant,
the agreement of the $R_{\rm{Ho}}/R_{25}$ ratio supports a similar
warp pattern in the Milky Way as observed for Briggs' spiral galaxies.

\begin{figure}
\begin{center}
\includegraphics[width=120mm]{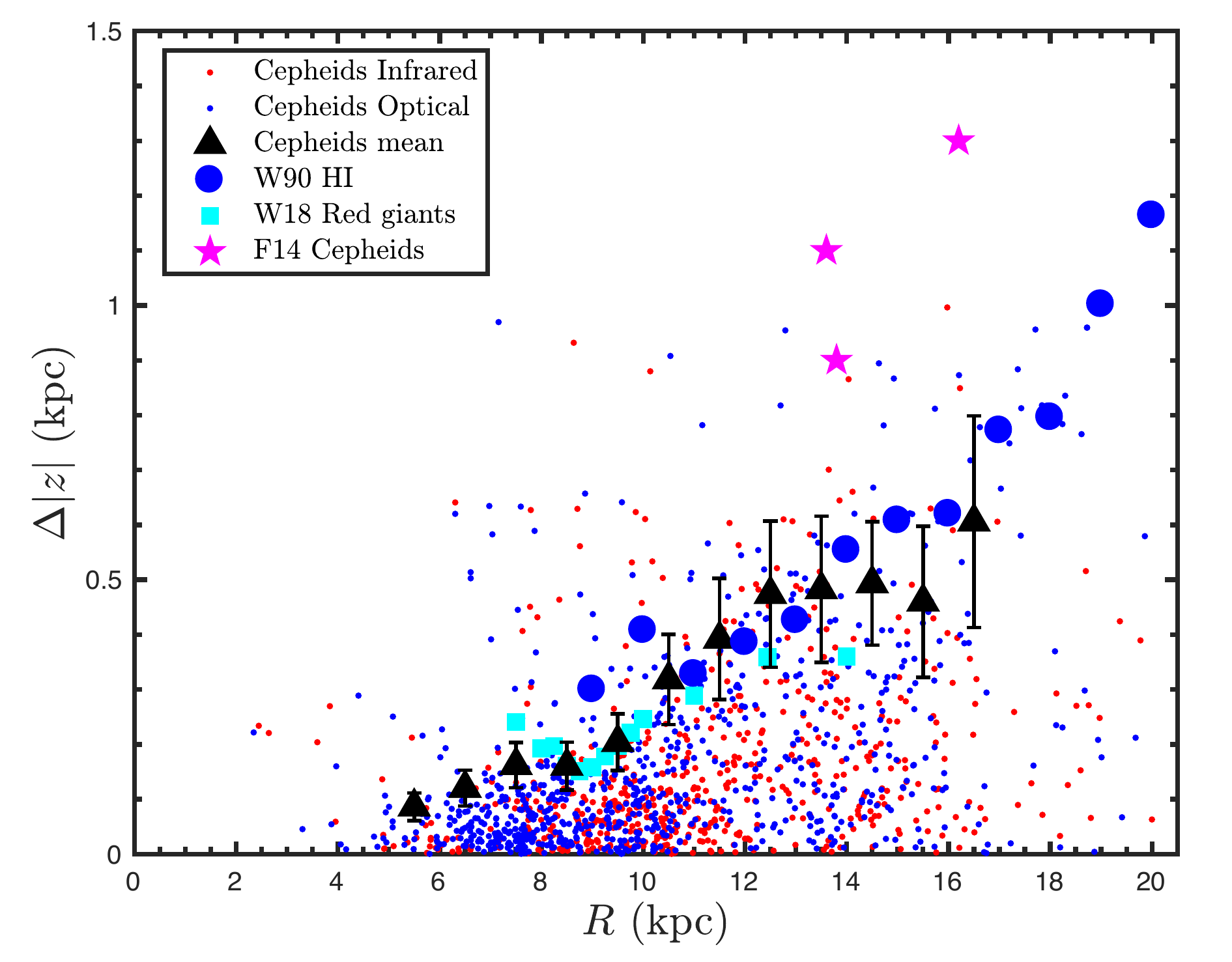}
\caption{\label{f4}{\bf Representation of the Milky Way's flare as
    traced by Cepheids.} Again, red and blue dots are Cepheids found
  in infrared (WISE) and optical passbands, respectively. $\Delta |z|$
  denotes the scale height of the flare, which is the difference in
  $z$ height between the Cepheids and the warp model. The black
  triangles denote the height of the flare in bins of 1 kpc
  Galactocentric radius. They agree well with the gas flare
  (W90\cite{Wouterloot90}: see the large blue dots) and the red
  giants' flare (W18\cite{Wangh18}: see the cyan squares) in the
  region of overlap. Three of the five Cepheids (close to the
    plane) previously found in the flare (F14\cite{Feast14}) are
  shown as magenta stars.}
\end{center}
\end{figure}

\begin{figure}
\begin{center}
\includegraphics[width=120mm]{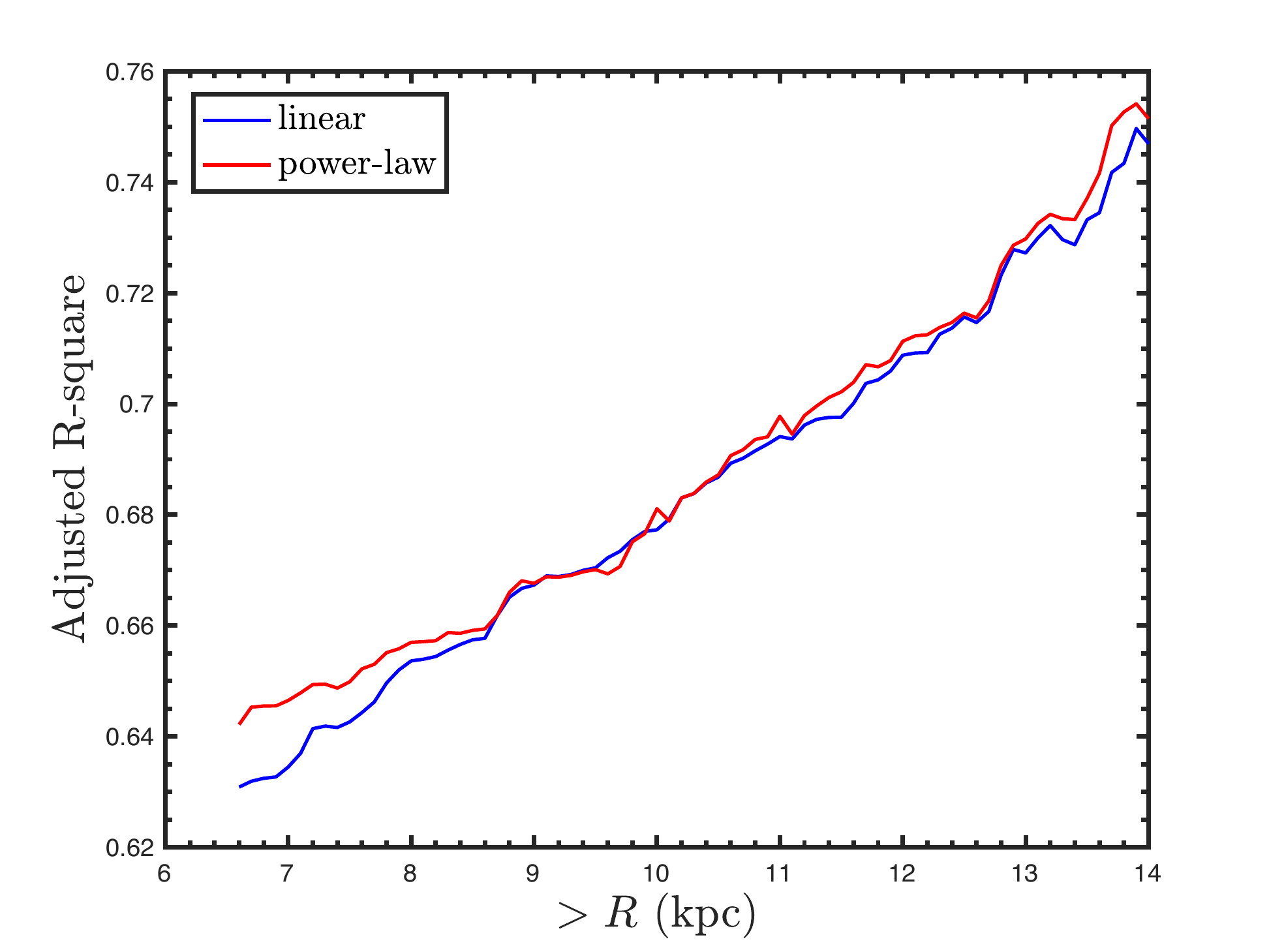}
\caption{\label{extend1} {\bf Comparison of the power-law and linear
    warp models.} Adjusted $R^2$ of the model fits for Cepheids at
  different Galactocentric radial ranges. Red and blue lines denote
  the adjusted $R^2$ of the power-law and linear models.}
\end{center}
\end{figure}

Finally, we estimated the $z$-height residuals, $\Delta|z|$: see
Figure \ref{f4}. The clear flare seen in the $\Delta|z|$ residuals
confirms the high reliability of both the data and our warp model. To
quantify the parameters of the flare, we estimated the scaleheight
based on the top-10 percentile of Cepheids in 1 kpc bins. The Cepheid
flare agrees well with the H{\sc i} flare in the region of overlap. In
detail, the Cepheid flare is smoother in the inner disk, whereas the
H{\sc i} flare is better defined in the outer disk because of the
decreasing completeness of our Cepheid sample at those radii. Three of the
  five previously confirmed Cepheids\cite{Feast14} which are located
  behind the Galactic Center in the flare region are shown as magenta
  stars. If we assume that the flare morphology behind the Galactic
  Center is similar to that on the near side, these three Cepheids
  appear on the far end of the Cepheid (or gas) flare scaleheight at
  these radii.

\begin{methods}
\subsection{Cepheid sample selection and the optimal distance method}

The `WISE Cepheids' were detected based on the full five-year WISE
all-sky survey. They are affected by incompleteness for long
periods ($P > 10$ days) and in crowded regions in the inner disk
(because of the WISE observation model). The completeness of
  Cepheids in the WISE variables catalogue\cite{Chen18a} is
  approximately 80\% (with respect to the optical Cepheid
  catalogue\cite{Berdnikov08}) in the magnitude range of
  interest. Significant incompleteness of the Cepheids in the WISE
  variables catalogue occurs at long periods ($P>10$ days) and for low
  amplitudes (${{\rm Amp}_{W1}}<0.2$ mag). WISE is ineffective in
detecting overtone Cepheids, which are characterised by half the
amplitudes of fundamental Cepheids. For example, based on 9649
  classical Cepheids in the Magellanic Clouds\cite{Soszynsky15}, the
  mean $I$-band amplitudes of the fundamental-mode and first-overtone
  Cepheids are 0.47 mag and 0.20 mag for $P>2$ days (the period range
  where both types of Cepheids overlap). Statistically, the fraction
  of first-overtone Cepheids is only 3.4\% (8 out of 237) among the
  Magellanic Cloud Cepheids rediscovered in the WISE variables
  catalogue. Because of the even smaller number of photometric
  detections in the Milky Way, no known Galactic overtone Cepheids in
  the optical Cepheid Catalogue\cite{Berdnikov08} were rediscovered in
  the WISE catalogue.

 Our optical Cepheid sample is based on detections in optical
  passbands. The sample was compiled based on both Cepheid
  catalogues\cite{Fernie95, Berdnikov08} and variable star
  catalogues\cite{Pojmanski05, Samus17, Jayasinghe18, Heinze18,
    Clementini18}. WISE Cepheids were not double counted. Since not
all of these catalogues clearly separate fundamental-mode and overtone
Cepheids, the optical sample may be affected by distance problems
caused by unrecognised overtone Cepheids. Nevertheless, the dominant
contaminants are Type-II Cepheids, eclipsing binaries and rapidly
rotating stars. By virtue of accurate {\sl Gaia} parallaxes at
distances within 5 kpc, most of these contaminants have been excluded
(see below).

Distances to our selected Cepheids were determined using the infrared
multi-passband optimal distance method\cite{Chen18b}. We adopted the
2MASS $JHK_{\rm s}$, {\sl Spitzer Space Telescope}\cite{Churchwell09}
[3.6], [4.5], [5.8] and [8.0], and WISE\cite{Wright10} $W1$ and $W2$
filters. In each band, the distance modulus was estimated using ${\rm
  DM}_{\lambda}=\langle m_{\lambda} \rangle - M_{\lambda}
-A_{\lambda}$. Here, $\langle m_{\lambda} \rangle$ is the mean
apparent magnitude, $M_{\lambda}$ the absolute magnitude determined
from the Galactic Cepheid period--luminosity
relations\cite{Wang18,Chen17} and $A_{\lambda}$ the extinction given
by the infrared extinction law\cite{Chen18b} and $A_{K_{\rm s}}$. The
$K_{\rm s}$-band extinction was adjusted to achieve a weighted average
distance modulus with the smallest possible standard deviation. The
weights were based on the total uncertainties in ${\rm DM}_{\lambda}$,
which include the photometric error, deviations of single-epoch
magnitudes from the mean magnitude, as well as period and extinction
uncertainties. Weights were set to 0 in bands without detections. The
extinction uncertainty $\sigma_{\lambda,{\rm ext}}$ increases as
$A_{K_{\rm s}}$ increases, so the extinction uncertainty dominates the
weights for objects affected by higher extinction. In fact, this
method yields the optimal distance based on a balance of extinction
and photometric errors. The near-infrared distance is usually
  determined as ${\rm DM}=\langle m_{K_{\rm s}} \rangle-M_{K_{\rm
      s}}-R_{K_{\rm s}} \times E(H-K_{\rm s})$
  \cite{Matsunaga11,Dekany15}. If we adopt a weight of unity in $H,
  K_{\rm s}$ and 0 in any other band, both methods become identical.

For the full sample, the statistical error is the larger of the
propagated error and the internal fitting error of the optimal
distance. The propagated error includes photometric uncertainties,
deviations from the mean magnitude, the intrinsic scatter in the
period--luminosity relations and period uncertainties. For
  single-epoch 2MASS photometry, the mean uncertainties associated
  with conversion to mean magnitudes are 0.100, 0.082 and 0.076 mag in
  $JHK_{\rm s}$, respectively, if we adopt the average full amplitudes
  $0.345\pm0.091$, $0.286\pm0.087$ and $0.265\pm0.087$ mag,
  respectively, based on 275 fundamental-mode
  Cepheids\cite{Inno15}. Some Cepheids observed with the {\sl Spitzer
  Space Telescope} have observations obtained during two epochs; the
adopted error is 0.05 mag. The amplitude relations in near-infrared
bands are based on template light curves\cite{Inno15}. Mid-infrared
amplitudes are assumed to be no larger than those in the $K_{\rm s}$
band. If these uncertainties are independent in each band, the final,
propagated uncertainties are $\sigma_{1}=\sqrt{1/(\sum
  {1/{\sigma_\lambda}^2})}$. The intrinsic scatter in the
period--luminosity relations for different wavelengths is
notindependent; therefore, the scatter in the $W1$ band is adopted
here (0.082 mag).

As regards the systematic error, the main contributors are the zero
point of the period--luminosity relation and the choice of extinction
law. The zero-point uncertainty of the infrared period--luminosity
relation is around 0.033 mag\cite{Freedman12,Chen18b}. The uncertainty
in the near-infrared extinction law is the main contributor to the
distance error in the Galactic plane; it can be up to 15\% for heavily
obscured stars\cite{Matsunaga18}. We adopted the infrared extinction
law determined using Cepheids in the Galactic Center direction
  (see Table \ref{ext1}, first row). It is comparable to the disk's
extinction law based on red clump
stars\cite{Zasowski09,Xue16,Indebetouw05} if we adopt the same
near-infrared extinction index $\alpha$. Since $\alpha$ could be
  variable, we estimate the mean bias in the distance modulus for
  different infrared extinction laws (see Table \ref{ext1}). Note that
  if $A_{W1}/A_{K_{\rm s}}$ and $A_{W2}/A_{K_{\rm s}}$ are not
  available\cite{Zasowski09,Indebetouw05}, the relative extinction
  values pertaining to the nearby [3.6] and [4.5] bands are adopted.
Half of the difference in the distance modulus associated with
adopting either $\alpha=1.61$ \cite{Cardelli89} or $\alpha=2.05$ was
treated as the error in the extinction; the mean deviation was 0.046
mag. In all of these statistical and systematic uncertainties,
uncertainties caused by the intrinsic scatter in the
period--luminosity relations and the optimal distance fitting
dominate.

\begin{table}
\tiny
\vspace{-0.0in}
\begin{center}
\caption{\label{ext1}Adopted infrared extinction laws and possible biases affecting the distance modulus.}
\vspace{0.15in}
\begin{tabular}{lcccccccccc}
\hline
\hline                               
                 &      $\alpha$  &  $A_J/A_{K_{\rm s}}$  &  $A_H/A_{K_{\rm s}}$  &  $A_{W1}/A_{K_{\rm s}}$  &  $A_{W2}/A_{K_{\rm s}}$  &  $A_{[3.6]}/A_{K_{\rm s}}$ &  $A_{[4.5]}/A_{K_{\rm s}}$ &  $A_{[5.8]}/A_{K_{\rm s}}$ &  $A_{[8.0]}/A_{K_{\rm s}}$ & Bias (mag)\\
\hline                        
ref. 9      &     2.05   &  3.005     & 1.717    & 0.506  &  0.340 & 0.478   &   0.341  &  0.234  &  0.321  & 0        \\
                      &     1.61   &  2.438     & 1.501    & 0.657  &  0.551 & 0.626   &   0.549  &  0.489  &  0.519  & 0.092    \\
ref. 46       &     1.79   &  2.720     & 1.599    & 0.591  &  0.463 & 0.553   &   0.461  &  0.389  &  0.426  & 0.044    \\
ref. 45  &     1.66   &  2.660     & 1.545    &   &   & 0.553   &   0.451  &  0.334  &  0.372  & 0.042    \\
ref. 47&    1.66   &  2.50      & 1.54     &   &   & 0.560   &   0.430  &  0.430  &  0.430  & 0.045    \\             
\hline
\end{tabular}
\end{center}
\end{table}

\subsection{Exclusion of Contaminants}

The main contaminants, Type-II Cepheids and long-period contact
binaries, were excluded based on parallax determinations from {\sl
  Gaia} Data Release 2. Reliable parallaxes were selected by
  requiring $\varpi>0.2$ mas, $\sigma_\varpi/\varpi<0.2$ and $G<16$
  mag, where $\varpi$ and $\sigma_\varpi$ are the {\sl Gaia}
  parallaxes and their uncertainties, respectively, and G denotes Gaia G band magnitudes.  False Cepheids
  were excluded based on the large differences between parallaxes
  derived from the period--luminosity relation distances and the
  actual {\sl Gaia} parallaxes, $|\varpi_{\rm
    PL}-\varpi|>3\sigma_{\varpi_t}$. Here, $\sigma_{\varpi_t}$ is the
  square root sum of the parallax error and the photometric distance
  error. We did not correct for possible systematic offsets in the
  {\sl Gaia} parallaxes, since any such offset is small compared with
  the other uncertainties we need to deal with. Nevertheless, we
  tested implementation of a correction of $-46$
  $\mu$as\cite{Riess18}. The number of objects in our final sample
  only decreased by 29 (corresponding to Cepheids with parallax
  differences in the range 3--$5\sigma_{\varpi_t}$). This has a
  negligible influence on the resulting mean warp parameters and the
  LON: the difference associated with adopting the corrected or
  uncorrected parallaxes is less than 10\% of the statistical
  uncertainty. Type-II Cepheids are typically 2--3 mag fainter than
classical Cepheids for a given period, whereas long-period contact
binaries are at least 4 mag fainter. If Type-II Cepheids at a true
distance of 5 kpc were mistaken for classical Cepheids, distances of
12.5--20 kpc would be estimated, somewhat depending on the pulsation
period. Long-period contact binaries at 5 kpc would be placed at
distances in excess of 30 kpc if they were assumed to be classical
Cepheids. Since {\sl Gaia} parallaxes are reliable out to distances of
order 5 kpc, they can be used as independent distance tracers to
exclude contaminants. This thus ensures the integrity of our Cepheid
sample within approximately 15 kpc.

Type-II Cepheids at distances of 5--8 kpc and $z$ heights $|z|<0.8$
kpc ($R=15$--20 kpc, $|z|<2.0$ kpc if treated as Type-I Cepheids) may
not be umambiguously excluded based on the {\sl Gaia}
parallaxes. In young environments, the ratio of Type-II to Type-I
Cepheids is small, however. Thanks to the warp feature, this ratio
could be estimated. Since the warp is not obvious for $R<10$ kpc,
  we can assume a symmetrical distribution of Type-II Cepheids at
  positive and negative $z$ heights. If they are treated as Type-I
  Cepheids, half of the Type-II Cepheids would appear at $z=-z_{\rm
    w}$. In other words, a false warp will be produced by these
  remnant Type-II Cepheids. Based on this idea, the number of Cepheids
  located within 0.5 kpc in $z$ height of a false warp and 1.0 kpc
  away from the real warp are considered contaminations. The
percentage of contaminants is $2n(-z_{\rm w})/n(z_{\rm w})=8$\%, where
$n(z_{\rm w})$ and $n(-z_{\rm w})$ are the numbers of Cepheids located
in the real and the falsely negative warp in the raw sample,
respectively. The 4\% Type-II Cepheids in the falsely negative warp
were excluded by the selection cut, whereas another (negligible) 4\%
(5 objects) may remain mixed in with our final sample.

\subsection{Validation of the warp model}

We considered both linear and power-law models to model the
warp. Figure \ref{extend1} shows that the power-law model is better
than the linear model at radii up to $R=7$--9 kpc. For objects at
$R>9$ kpc, the two models are comparable. This means that the linear
model is not suitable for Cepheids at $R< 9$ kpc. Therefore, the
results for the linear model were determined using Cepheids at
Galactocentric distances greater than 9 kpc. MCMC simulations were
performed to verify the warp model and investigate correlations among
the parameters. In Figure \ref{extend2}, the Gaussian distributions
and the similar values validate the results of the nonlinear
least-squares method.

We converted $R\phi z$ to $R\phi\theta$ to investigate the warp model
in spherical coordinates and adopted $\psi=a\psi_{\rm w}(R-R_{\rm
  w})^b\sin(\phi-\phi_{\rm w})$\cite{Abedi14}, where $\psi$ is the
tilt angle. The $\phi$ values thus determined, both the mean value and
the corresponding values as a function of radius, based on this warp
model are almost the same as those of our above results. This means
that the spatial distortion caused by adopting spherical coordinates
is small. This can be understood based on two arguments. First, the
tilt angle of the Cepheid warp is small ($4^\circ$ at a distance of 17
kpc; see Figure \ref{f2}), so the spatial distortion is
negligible. Second, $\phi$ is almost independent of $R, z$ (see Figure
\ref{extend2}), so that adoption of $R, z$ or $R, \theta$ has little
influence on $\phi$.

To test whether or not the $m=2$ warp model is realistic, we
rearranged the model to read $z_{\rm w}=W_0+W_1(R-R_{\rm
  w})\sin(\phi-\phi_{\rm w})+W_2(R-R_{\rm W2})\sin(2\phi-\phi_{\rm
  W2})$\cite{Levine06}. $W_0, W_1, W_2$ are the $z$ amplitudes of the
$m=0, 1, 2$ modes, respectively, and $\phi_{\rm w}, \phi_{\rm W2}$ are
the LON angles for the $m=1, 2$ modes. For convenience, $R_{\rm W2}$
was adopted as the sample's minimum Galactocentric distance. $W_2$ is
around $0.01\pm0.01$ and $0.02\pm0.02$ kpc kpc$^{-1}$ for Cepheids in
the range $R<10, 10<R<12$ kpc, which means that the $m=2$ warp model
is not obvious at $R<12$ kpc. Analysis of 146 Cepheids at $R>15$ kpc
shows an obvious $m=2$ amplitude, $W_2=0.14\pm0.03$ kpc kpc$^{-1}$,
similar to that of the H{\sc i} model, $W_2 ({\rm H\sc I})=0.12$ kpc
kpc$^{-1}$, whereas $\phi_{\rm W2}$ is rather different. For the gas
model, the line of maxima for the $m=2$ mode is roughly aligned with
the lines of maxima of the $m=1$ mode, which is different from that of
the Cepheids: the lines of maxima for the two modes deviate by
approximately $45^\circ$. The best values from the MCMC simulation are
shown in Figure \ref{extend3}.

\subsection{Validation of the warp's LON}

To verify the apparent precession trend of the warp's LON shown in
Figure \ref{f3}, different conditions which would affect the result
are considered. The LON in each radial bin is first tested based on
MCMC simulations with fixed and free $a, b, R_{\rm w}$ parameters. The
50, 16 and 84 percentiles in the probability distribution are adopted
as the median value and the corresponding errors. A comparison with
the results of our nonlinear least-squares fitting method is shown in
Figure \ref{extend4}a, b. Since the Cepheid sample was selected by
imposing a limiting distance accuracy of 5\%, a test was done using a
different sample containing Cepheids with distance accuracies better
than 10\%. The warp's LONs in different radial bins were again
analysed using MCMC simulations. The trend is shown in Figure
\ref{extend4}c. The excellent agreement of LON trends in Figure
\ref{extend4}a, b, c means that the LON is robust among different
methods and samples.

The effect of the propagation of distance uncertainties was quantified
by means of Monte Carlo simulations. In addition to the statistical
uncertainties, we also simulated the deviations caused by inclusion of
10\% overtone Cepheids and of 2.6\% systematic distance
uncertainties. Based on 2000 realisations, the mean LONs and their
standard deviations are shown in Figure \ref{extend4}d. Another
possible systematic effect may be caused by the unequal distributions
of Cepheids in the northern and southern warps, or by their spatial
clumpiness. Since the number of Cepheids in the northern warp is half
that in the southern warp, we resampled the southern warp to consider
equal numbers. To avoid Cepheid clumpiness, we did not include
Cepheids at $|\phi|<10^\circ$ in our tests. The sample was randomly
selected 1000 times; the mean LONs and their standard deviation are
plotted in Figure \ref{extend4}e. Figure \ref{extend4} shows that all
LON trends agree well, which serves as strong validation of the
precession trend. The final uncertainties in the LONs are based on the
largest of the statistical uncertainties and the systematic
deviations.

\subsection{Kinematics of the Cepheid warp}

As shown in the kinematic map based on upper main-sequence stars and
giants\cite{Poggio18}, the maximum median value $v_z$ is around 7.5 km
s$^{-1}$. Indeed, stars around the warp LONs have higher absolute
velocities, $v_z$. Investigation of the LON of the kinematic warp
requires even higher accuracies for both distance and velocity
measurements. Limited by the larger mean uncertainty $\langle
\sigma_{v_z} \rangle = 4.2$ km s$^{-1}$, only Cepheids with
uncertainties $\sigma_{v_z}<5$ km s$^{-1}$ appear in Figure
\ref{extend5}. We therefore only use the kinematic map as an
additional tool to characterise the spatial warp. The blue and red
data points are Cepheids with obviously positive and negative $v_z$,
respectively. They are indeed reliable given their high
signal-to-noise ratios $>$ 3. The red dots representing the
  possible kinematic LONs evidently confirm a tilted LON and agree
  well with the spatial LONs (considering the prevailing
  uncertainties). This tilted LON traced by Cepheids is also
  consistent with that traced by upper main-sequence stars and
  giants\cite{Poggio18}. In addition, more negative $v_z$ Cepheids
are located around and beyond the maximum amplitude direction of the
southern warp, which confirms that the orientation of the LON is on
the left-hand side of the solar direction, $\phi>0$.

\end{methods}

%% Here is the endmatter stuff: Supplementary Info, etc.
%% Use \item's to separate, default label is "Acknowledgements"

\begin{addendum}
 \item [Acknowledgements] We are grateful for research support from
   the National Key Basic Research Program of China 2014CB845700. This
   work is also supported by the National Natural Science Foundation
   of China through grants U1631102, 11373010 and 11633005, the
   Initiative Postdocs Support Program (No. BX201600002), the China
   Postdoctoral Science Foundation (grant 2017M610998) and the
   National Key Research and Development Program of China (grant
   2017YFA0402702).
   
 \item [Author contributions] X.C. contributed to the project
   planning, data preparation and analysis, modeling, simulations and
   writing of the final paper. S.W. contributed to the data analysis and
   writing of the paper. L.D. contributed to project planning and research
   support. R.d.G. engaged in detailed scientific discussions and
   contributed to writing of the paper and final editing. C.L. contributed
   to the exploration of the warp's precession. H.T. contributed to
   implementation of the techniques used for the modeling and
   simulations. All authors reviewed and commented on the manuscript.
   
 \item[Competing Interests] The authors declare that they have no
   competing financial interests.
 \item[Correspondence] Correspondence and requests for materials
   should be addressed to Xiaodian Chen (email:
   chenxiaodian@nao.cas.cn), Shu Wang (email: shuwang@pku.edu.cn) or
   Licai Deng (email: licai@bao.ac.cn).
% \item[Additional information]
% Extended data is published with the paper at ...
\end{addendum}

%%
%% TABLES
%%
%% If there are any tables, put them here.
%%

\begin{figure}
\begin{center}
\includegraphics[width=\textwidth]{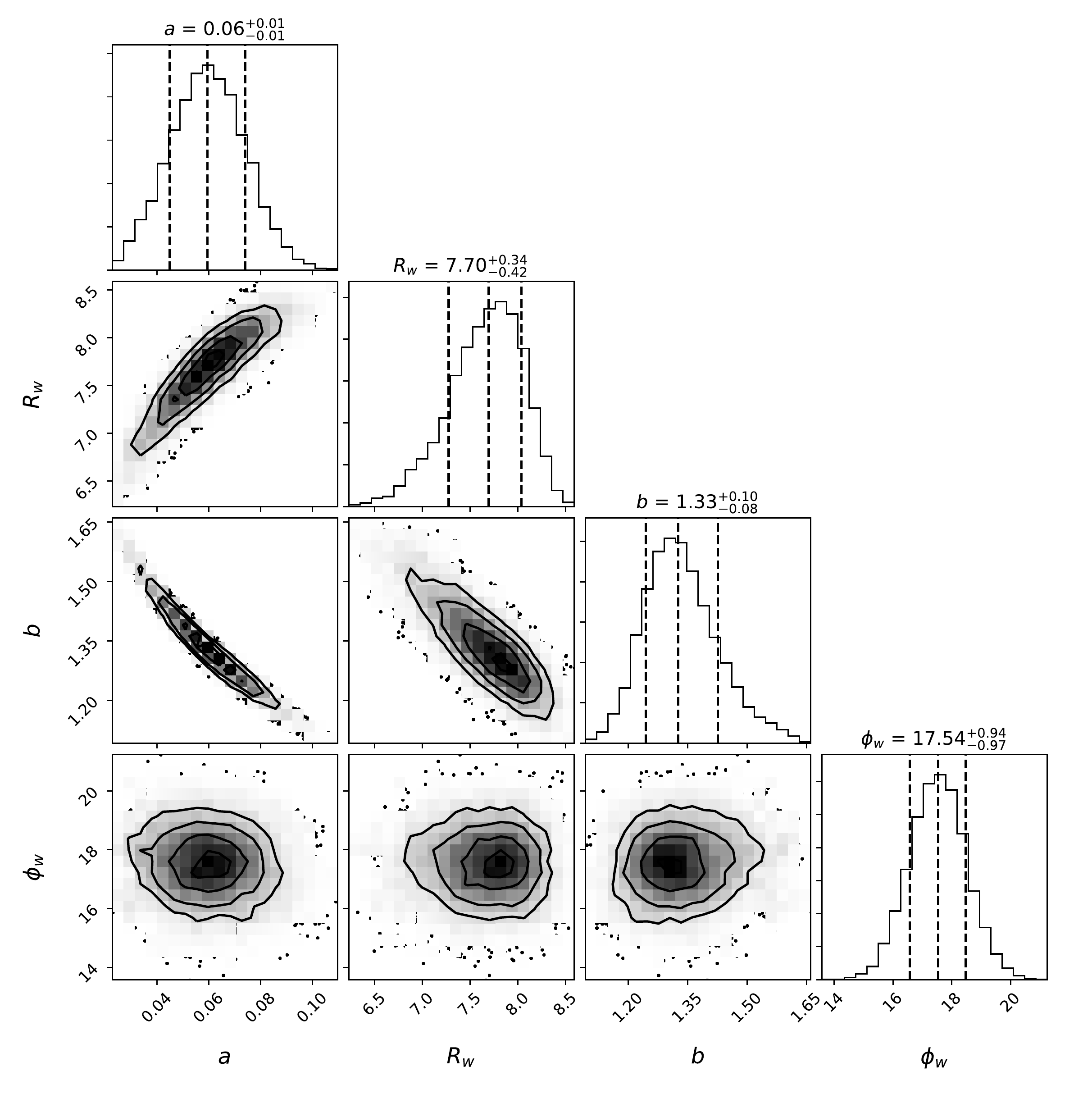}
\caption{\label{extend2} {\bf Probability distribution of the
    parameters in the $z_{\rm w}=a(R-R_{\rm w})^b\sin(\phi-\phi_{\rm
      w})$ warp models determined based on our MCMC simulation.} The
  median value and the 16 and 84 percentile probabilities are
  indicated.}
\end{center}
\end{figure}

\begin{figure}
\begin{center}
\includegraphics[width=\textwidth]{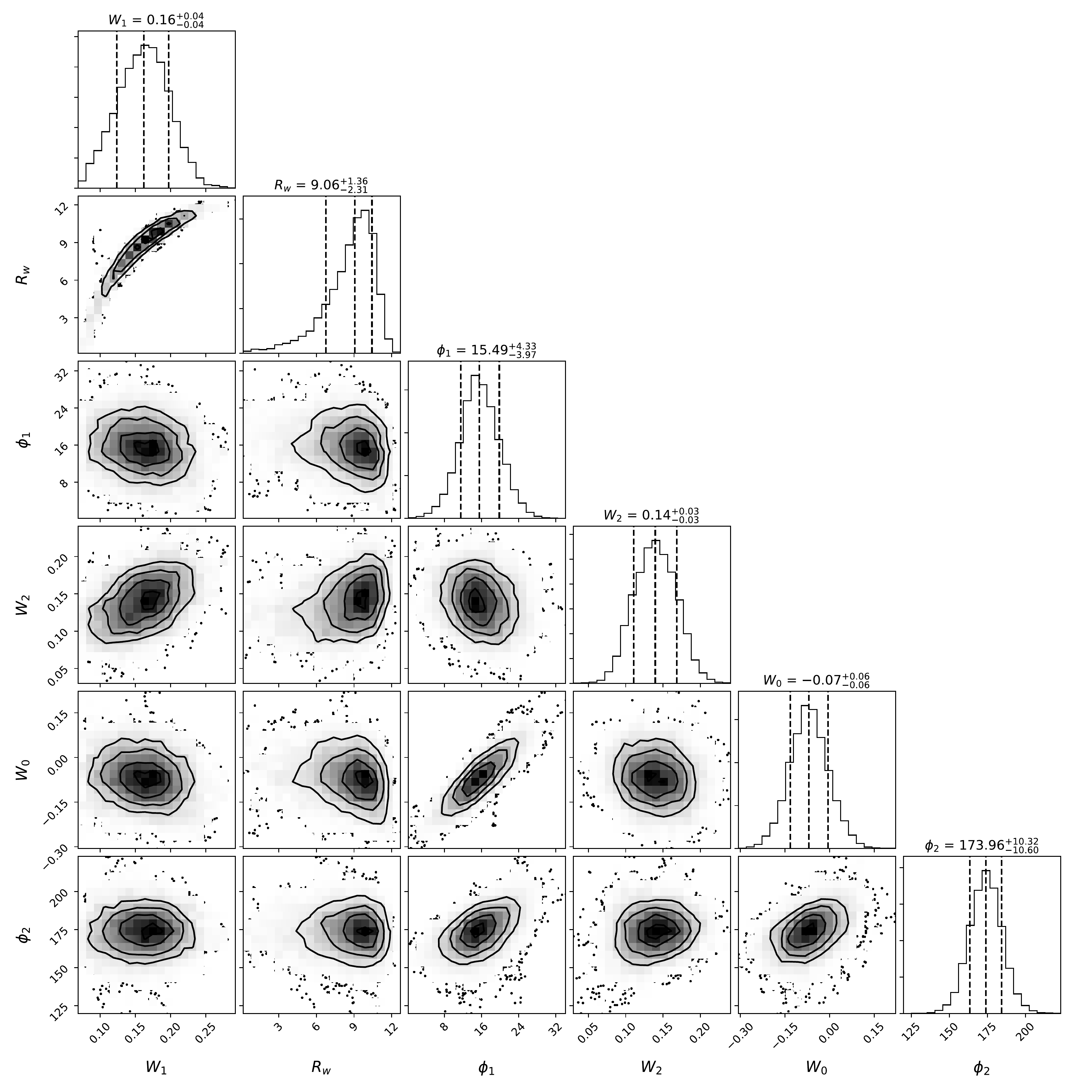}
\caption{\label{extend3} {\bf Probability distribution of the
    parameters in the $z_{\rm w}=W_0+W_1(R-R_{\rm
      w})\sin(\phi-\phi_{\rm w})+W_2(R-15)\sin(2\phi-\phi_{\rm w2})$
    warp models for the $m=0, 1, 2$ modes based on 146 Cepheids at
    $R>15$ kpc.} The median value and the 16 and 84 percentile
  probabilities are indicated.}
\end{center}
\end{figure}

\begin{figure}
\begin{center}
\includegraphics[width=120mm]{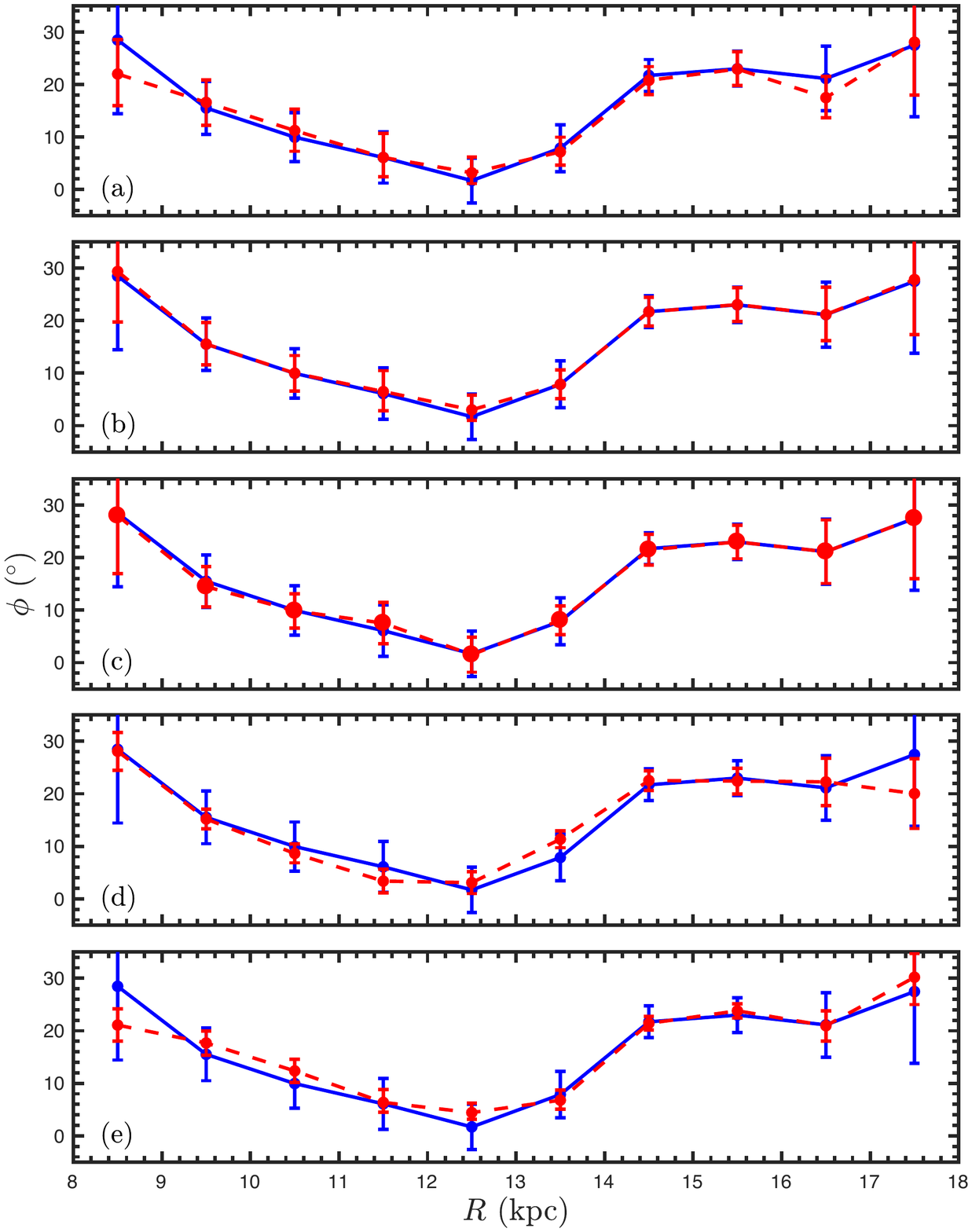}
\caption{\label{extend4}{\bf Validation of the warp's LON for
    different conditions.} Blue dots denote the LONs determined based
  on application of the nonlinear least-squares method to the warp
  model (identical to the blue dots in Figure \ref{f3}). The error
  bars include all systematic and statistical uncertainties. Red dots
  in each panel denote LONs determined under different conditions. a:
  MCMC simulation with free parameters; b: MCMC simulation with fixed
  parameters; c: Cepheids selected based on a 10\% accuracy cut in
  distances; d: error propagation considered in the Monte Carlo
  simulations; e: resampling test to consider equal numbers of
  Cepheids in the northern and southern warps.}
\end{center}
\end{figure}

\begin{figure}
\begin{center}
\includegraphics[width=180mm]{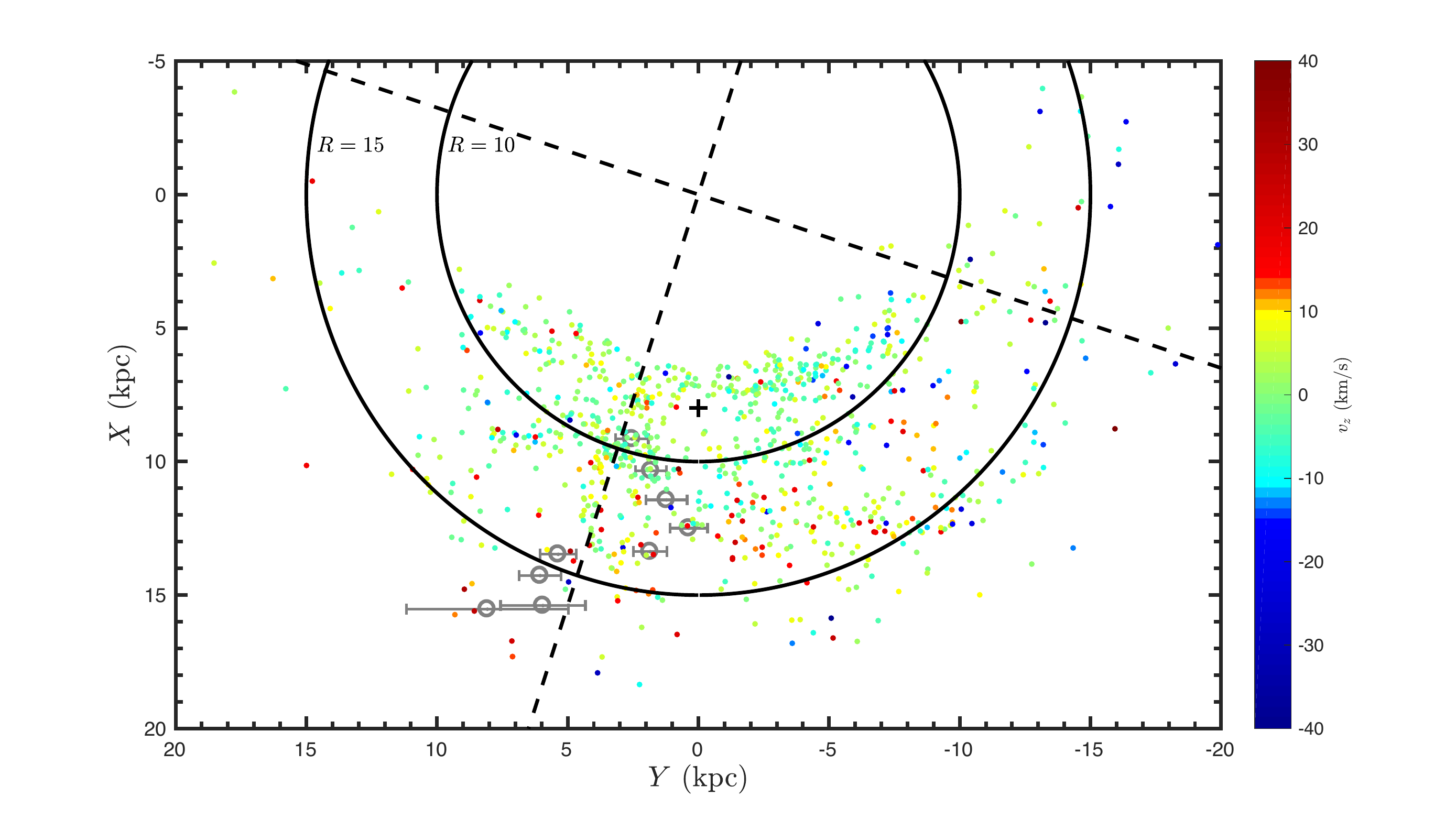}
\caption{\label{extend5} {\bf $v_z$ versus $XY$ map traced by
    Cepheids.} Cepheids with large positive $v_z$ are shown in red,
  whereas their negative counterparts are shown in blue. The large
  black circles denote $R=10, 15$ kpc and black dashed lines denote
  the LON line (close to Sun) and the warp's maxima line. The Sun is
  shown as the black plus sign and the spatial LONs are represented by
  grey circles.}
\end{center}
\end{figure}


\begin{thebibliography}{}
\bibitem[1]{Levine06} Levine, E. S., Blitz, L. \& Heiles, C. The
  vertical structure of the outer Milky Way H{\sc i}
  disk. Astrophys. J. 643, 881--896 (2006).
\bibitem[2]{Kalberla07} Kalberla, P. M. W., Dedes, L., Kerp, J. \&
  Haud, U. Dark matter in the Milky Way. II. The H{\sc i} gas
  distribution as a tracer of the gravitational
  potential. Astron. Astrophys. 469, 511--527 (2007).
\bibitem[3]{Wang18} Wang, S., Chen, X., de Grijs, R. \& Deng, L. The
  Near-infrared Optimal Distances Method Applied to Galactic Classical
  Cepheids Tightly Constrains Mid-infrared Period--Luminosity
  Relations. Astrophys. J. 852, 78 (2018).
\bibitem[4]{Briggs90} Briggs, F. H. Rules of behavior for galactic
  warps. Astrophys. J. 352, 15--29 (1990).
\bibitem[5]{Shen06} Shen, J., \& Sellwood, J. A. Galactic warps
  induced by cosmic infall. Mon. Not. R. Astron. Soc. 370, 2--14
  (2006).
\bibitem[6]{Liu17} Liu, C., Xu, Y., Wang, H., \& Wan, J. Rediscovering
  the Galactic outer disk with LAMOST data. Proc. Int. Astron. Union Vol. 13 (eds Chiappini, C., Minchev, I., Starkenberg, E. \& Valentini, M.) 109–115 (International Astronomical Union, 2018).
\bibitem[7]{Wangh18} Wang, H., Liu, C., Xu, Y., Wan, J., \& Deng,
  L. Mapping the Milky Way with LAMOST- III. Complicated spatial
  structure in the outer disc. Mon. Not. R. Astron. Soc. 478,
  3367--3379 (2018).
\bibitem[8]{Chen18a} Chen, X., Wang, S., Deng, L., de Grijs, R. \&
  Ming, Y. Wide-field Infrared Survey Explorer (WISE) Catalog of
  Periodic Variable Stars. Astrophys. J. Suppl. Ser., 273, 28 (2018).
\bibitem[9]{Chen18b} Chen, X., Wang, S., Deng, L. \& de Grijs, R. An
  Extremely Low Mid-infrared Extinction Law toward the Galactic Center
  and 4\% Distance Precision to 55 Classical
  Cepheids. Astrophys. J. 859, 137 (2018).
\bibitem[10]{Gaia18} Gaia Collaboration. Gaia Data Release 2. Summary of the contents
  and survey properties. Astron. Astrophys. 616, A1 (2018).
\bibitem[11]{Haario06} Haario, H., Laine, M., Mira, A., \& Saksman,
  E., DRAM: Efficient adaptive MCMC, Statistics and Computing 16,
  339--354 (2006).
\bibitem[12]{Abedi14} Abedi, H., Mateu, C., Aguilar, L. A. et
  al. Characterizing the Galactic warp with Gaia - I. The tilted ring
  model with a twist. Mon. Not. R. Astron. Soc. 442, 3627--3642
  (2014).
\bibitem[13]{Drimmel01} Drimmel, R., \& Spergel,
  D. N. Three-dimensional Structure of the Milky Way Disk: The
  Distribution of Stars and Dust beyond 0.35 $R_0$. Astrophys. J. 556,
  181--202 (2018).
\bibitem[14]{Lopez02} L\'{o}pez-Corredoira, M., Cabrera-Lavers, A.,
  Garz\'{o}n, F., \& Hammersley, P. L. Old stellar Galactic disc in
  near-plane regions according to 2MASS: Scales, cut-off, flare and
  warp. Astron. Astrophys, 394, 883--899 (2002).
\bibitem[15]{Yusifov04} Yusifov, I. Pulsars and the warp of the Galaxy. In The Magnetized Interstellar Medium: Proc. Conference held in Antalya, Turkey (eds Uyaniker, B., Reich, W. \& Wielebinski, R.) 165–169 (Copernicus GmbH, 2004).
\bibitem[16]{Skrutskie06} Skrutskie, M. F., Cutri, R. M., Stiening,
  R., et al. The Two Micron All Sky Survey (2MASS). Astron. J. 131,
  1163--1183 (2006).
\bibitem[17]{Amores17} Amores, E. B., Robin, A. C., Reyl\'e,
  C. Evolution over time of the Milky Way's disc
  shape. Astron. Astrophys. 602, A67 (2017).
\bibitem[18]{Poggio18} Poggio, E., Drimmel, R., Lattanzi, M. G. et
  al. The Galactic warp revealed by Gaia DR2
  kinematics. Mon. Not. R. Astron. Soc. 481. L21--L25 (2018).
\bibitem[19]{Burton88} Burton, W. B. in Galactic and Extragalactic Radio Astronomy 2nd edn (eds Verschuur, G. \& Kellermann, K.) 295–358 (Springer-Verlag, Berlin and New York, 1988).
\bibitem[20]{Momany06} Momany, Y., Zaggia, S., Gilmore, G. et
  al. Outer structure of the Galactic warp and flare: explaining the
  Canis Major over-density. Astron. Astrophys. 451, 515--538 (2006).
 \bibitem[21]{Bland-Hawthorn16} Bland-Hawthorn, J., \& Gerhard, O. The
   Galaxy in Context: Structural, Kinematic, and Integrated
   Properties. ARA\&A, 54, 529--596 (2016).
\bibitem[22]{deGrijs16} de Grijs, R., \& Bono, G. Clustering of Local
  Group Distances: Publication Bias or Correlated Measurements?
  IV. The Galactic Center. Astrophys. J. Suppl. Ser. 227, 5 (2016).
\bibitem[23]{Freedman81} Freedman, D., Diaconis, P., On the histogram
  as a density estimator: L2 theory. Probability Theory and Related
  Fields. Heidelberg: Springer Berlin. 57 (4): 453--476 (1981).
\bibitem[24]{Reid14} Reid, M. J., Menten, K. M., Brunthaler, A. et
  al. Trigonometric Parallaxes of High Mass Star Forming Regions: The
  Structure and Kinematics of the Milky Way. Astrophys. J. 783, 130
  (2014).
\bibitem[25]{Schonrich10} Schonrich R., Binney J., \& Dehnen W., Local
  kinematics and the local standard of
  rest. Mon. Not. R. Astron. Soc., 403, 1829--1833 (2010).
\bibitem[26]{Antoja18} Antoja, T., Helmi, A., Romero-Gomez, M. et
  al. A dynamically young and perturbed Milky Way disk. Nature. 561,
  360--362 (2018).
\bibitem[27]{Freeman70} Freeman, K. C. On the Disks of Spiral and S0
  Galaxies. Astrophys. J. 160, 811--830 (1970).
\bibitem[28]{Feast14} Feast, M. W., Menzies, J. W., Matsunaga, N.,
  Whitelock, P. A. Cepheid variables in the flared outer disk of our
  galaxy. Nature 509, 342--344 (2014).
\bibitem[29]{Wouterloot90} Wouterloot, J. G. A., Brand, J., Burton,
  W. B., \& Kwee, K. K. IRAS sources beyond the solar circle. II.
  Distribution in the Galactic warp. Astron. Astrophys. 230, 21--36
  (1990).
\bibitem[30]{Fernie95} Fernie, J. D., Evans, N. R., Beattie, B., \&
  Seager, S. A Database of Galactic Classical Cepheids. IBVS. 4148, 1
  (1995).
\bibitem[31]{Berdnikov08} Berdnikov, L. N. VizieR Online Data Catalog:
  Photoelectric observations of Cepheids in $UBV(RI)_c$ II/285 http://vizier.cfa.harvard.edu/viz-bin/VizieR?-source=II/285 (2008).
\bibitem[32]{Soszynsky15} Soszy\'nski, I., Udalski, A., Szyma\'nski,
  M. K.  et al. The OGLE Collection of Variable Stars. Classical
  Cepheids in the Magellanic System. Acta Astronomica, 65, 297--312
  (2015).  Photoelectric observations of Cepheids in $UBV(RI)_c$
  VizieR On-line Data Catalog: II/285 (2008).
\bibitem[33]{Pojmanski05} Pojmanski, G., Pilecki, B., \& Szczygiel,
  D. The All Sky Automated Survey. Catalog of Variable
  Stars. V. Declinations 0$^\circ$ -- +28$^\circ$ of the Northern
  Hemisphere. Acta Astronomica. 55, 275--301 (2005).
\bibitem[34]{Samus17} Samus, N. N., Kazarovets, E. V., Durlevich,
  O. V., Kireeva, N. N., \& Pastukhova, E. N.  General catalogue of
  variable stars: Version GCVS 5.1. Astron. Rep. 61, 80--88 (2017)
\bibitem[35]{Jayasinghe18} Jayasinghe, T., Kochanek, C. S., Stanek,
  K. Z et al. The ASAS-SN catalogue of variable stars. I. The
  Serendipitous Survey. Mon. Not. R. Astron. Soc. 477, 3145--3163
  (2018).
\bibitem[36]{Heinze18} Heinze, A. N., Tonry, J. L., Denneau, L., et
  al. A first catalog of variable stars measured by the Asteroid Terrestrial-impact Last Alert System (ATLAS). Astron. J. 156, 241 (2018).
\bibitem[37]{Clementini18} Clementini, G., Ripepi, V., Molinaro, R. et
  al. Gaia Data Release 2: Specific characterisation and validation of
  all-sky Cepheids and RR Lyrae
  stars. Astron. Astrophys. (in the press). arXiv:1805.02079 (2018)
\bibitem[38]{Churchwell09} Churchwell, E., Babler, B. L., Meade,
  M. R., et al. The Spitzer/GLIMPSE Surveys: A New View of the Milky
  Way. Publ. Astron. Soc. Pac. 121, 213--230 (2009).
\bibitem[39]{Wright10} Wright, E. L., Eisenhardt, P. R. M., Mainzer,
  A. K., et al. The Wide-field Infrared Survey Explorer (WISE):
  Mission Description and Initial On-orbit
  Performance. Astron. J. 140, 1868--1881 (2010).
\bibitem[40]{Chen17} Chen, X., de Grijs, R., \& Deng, L. New open
  cluster Cepheids in the VVV survey tightly constrain near-infrared
  period--luminosity relations. Mon. Not. R. Astron. Soc. 464,
  1119--1126 (2017).
\bibitem[41]{Matsunaga11} Matsunaga, N., Kawadu, T., Nishiyama, S. et
  al. Three classical Cepheid variable stars in the nuclear bulge of
  the Milky Way. Nature. 477, 188--190 (2011).
\bibitem[42]{Dekany15} D\'ek\'any, I., Minniti, D., Majaess, D. et
  al. The VVV Survey Reveals Classical Cepheids Tracing a Young and
  Thin Stellar Disk across the Galaxy’s Bulge. Astrophys. J. 812. L29
  (2015).
\bibitem[43]{Inno15} Inno, L., Matsunaga, N., Romaniello, M. et
  al. New NIR light-curve templates for classical
  Cepheids. Astron. Astrophys. 576, 30 (2015).
\bibitem[44]{Freedman12} Freedman, W. L., Madore, B. F., Scowcroft,
  V. et al. Carnegie Hubble Program: A Mid-infrared Calibration of the
  Hubble Constant. Astrophys. J. 758, 24 (2012).
\bibitem[45]{Matsunaga18} Matsunaga, N., Bono, G., Chen, X. et
  al. Impact of Distance Determinations on Galactic
  Structure. I. Young and Intermediate-Age Tracers. Space Science
  Reviews. 214, 74 (2018).
\bibitem[46]{Zasowski09}Zasowski, G., Majewski, S. R., Indebetouw,
  R. et al. Lifting the Dusty Veil with Near- and Mid-Infrared
  Photometry. II. A Large-Scale Study of the Galactic Infrared
  Extinction Law. Astrophys. J. 707, 510--523 (2009).
\bibitem[47]{Xue16}Xue, M., Jiang, B. W., Gao, J. et al. A Precise
  Determination of the Mid-infrared Interstellar Extinction Law Based
  on the APOGEE Spectroscopic Survey. Astrophys. J. Suppl. Ser. 224,
  23 (2016).
\bibitem[48]{Indebetouw05}Indebetouw, R., Mathis, J. S., Babler,
  B. L. et al. The Wavelength Dependence of Interstellar Extinction
  from 1.25 to 8.0 $\mu$m Using GLIMPSE Data. Astrophys. J. 619,
  931--938 (2005).
\bibitem[49]{Cardelli89} Cardelli, J. A., Clayton, G. C., \& Mathis,
  J. S., The relationship between infrared, optical, and ultraviolet
  extinction. Astrophys. J. 345, 245--256 (1989).
\bibitem[50]{Riess18} Riess, A., Casertano, S., Yuan, W. et al. Milky
  Way Cepheid Standards for Measuring Cosmic Distances and Application
  to Gaia DR2: Implications for the Hubble
  Constant. Astrophys. J. 861, 126 (2018).

\end{thebibliography}
\end{document}